\documentclass[twocolumn,aps,amssymb,showpacs,superscriptaddress,secnumarabic]{revtex4}
\usepackage{graphicx}
\usepackage{amsmath}
\usepackage{mathtools}
\usepackage{chngcntr}

\begin{document}
\date{\today}
\title {Reaction-Diffusion Theory in the Presence of an Attractive Harmonic Potential}

\author{K. Spendier} 
\email{kspendie@uccs.edu} 
\affiliation{BioFrontiers Center, University of Colorado at Colorado Springs, Colorado Springs, Colorado 80918}
\affiliation{Department of Physics and Energy Science, University of Colorado at Colorado Springs, Colorado Springs, Colorado 80918}

\author{S. Sugaya}
\email{satomi@unm.edu}
\affiliation{Consortium of the Americas for Interdisciplinary Science, University of New Mexico, Albuquerque, New Mexico 87131}
\affiliation{Department of Physics and Astronomy, University of New Mexico, Albuquerque, New Mexico 87131}

\author{V. M. Kenkre} 
\email{kenkre@unm.edu} 
\affiliation{Consortium of the Americas for Interdisciplinary Science, University of New Mexico, Albuquerque, New Mexico 87131}
\affiliation{Department of Physics and Astronomy, University of New Mexico, Albuquerque, New Mexico 87131}
\begin{abstract}
Problems involving the capture of a moving entity by a trap occur in a variety of physical situations, the moving entity being an electron, an excitation, an atom, a molecule, a biological object such as a receptor cluster, a cell, or even an animal such as a mouse carrying an epidemic. Theoretical considerations have almost always assumed that the particle motion is translationally invariant. We study here the case when that assumption is relaxed, in that the particle is additionally subjected to a harmonic potential. This tethering to a center modifies the reaction-diffusion phenomenon. Using a Smoluchowski equation to describe the system, we carry out a study which is explicit in 1 dimension but can be easily extended for  arbitrary dimensions. Interesting features emerge depending on the relative location of the trap, the attractive center and the initial placement of the diffusing particle.
\end{abstract}

\maketitle

\section{Introduction and the Smoluchowski Equation}

Reaction-diffusion problems of diffusing entities are of great interest in a broad range of physical systems in physics, chemistry, and biology, and have been typically studied under the assumption that, in the absence of the the reaction phenomenon, the motion is translationally invariant~\cite{Rice:1985, WILEMSKI:1973:ID221, Katja3, KENKRE:1983, REDNER:2001:ID188, KENKRE:1983:ID198, SZABO:1984:ID171, ABRAMSON:1995:ID169, KSinJPC}. Our interest in the present paper is to extend these studies fundamentally by going beyond that assumption. We analyze systems in which a particle diffuses in a harmonic potential and  undergoes capture at a given rate when it arrives at a fixed trap. The diffusion means that the particle performs a random walk. The attraction to the potential center means that a tendency to be tethered to a fixed point is present in addition. The location of the trap is generally arbitrary. e.g., not necessarily at the attractive center; this feature introduces interesting variety in the 
consequences of the reaction-diffusion phenomenon. 

Systems characterized by the above features occur often in nature. Molecular forces confine moving entities in various physical and chemical situations in sensitized luminescence and photosynthesis. Funneling phenomena attract excitations in photosynthetic antennae and molecular crystals and aggregates. Particle diffusion in a harmonic field occurs also in biophysical studies on DNA stretching with optical tweezers~\cite{opttweezers, opttweezers2}. Yet another relevant area is electrostatic steering in enzyme ligand binding~\cite{Enzyme1,Enzyme2}. To bind at an enzymes active site, a ligand must diffuse or be transported to the enzyme surface, and, if the binding site is buried, the ligand must diffuse through the protein to reach it. Enhancement of this diffusion can be achieved by attractive electrostatic interactions between the substrate and the protein binding site. On a more macroscopic scale, animals feel a driving force pointing towards their \emph{nest}~\cite{EBELING:1999:ID276} with the consequent emergence of home ranges~\cite{Giuggioli:2006}. Transmission of infection in terrains where infected animals (such as rodents) interact with susceptible ones under the action of space confinement provides a related and more complex area of study.

Depending on the relative position of the attractive center and the trap, a particle placed initially  at some location may be affected \emph{either} favorably \emph{or} unfavorably during different stages of its motion as far as the efficiency of the trapping phenomenon is concerned. In a translationally invariant system all that is relevant is the initial distance between the particle and reactive site, the manner of motion that occurs in between, and the rate at which the reaction occurs. The problem is rendered considerably richer in the presence of an attractive potential as we shall see below.

Our study of the literature has uncovered no major previous advances in reaction-diffusion theory in the presence of a potential. Of the two relevant articles we have found, refs. \cite{Drazer} and \cite{BAGCHI:1983:ID275}, the former has no position-dependence in the capture which is trivially represented via a constant term, the emphasis being on anomalous diffusion, and in the latter, only perfect absorption is treated and that only for a centrally located trap, which, as we shall show below, results in a relatively featureless case. Thus, a general understanding of the situation is not available in the literature. This has motivated us to undertake the present investigation. 

The outline of the paper is as follows. In the rest of this section we specify the Smoluchowski equation as the basic equation of motion that we start with, in the absence of capture. In Section 2, we show how the survival probability depends on the propagators of the equation of motion in general, and give explicit expressions for the propagators of the Smoluchowski equation. The combination of these two results is the point of departure of our analysis in the subsequent sections in the paper. Section 3 treats the simple case of a centrally placed trap for which we obtain an \emph{analytic} solution in the Laplace domain, involving the Whittaker function. Analytic inversion is possible for perfect absorption but finite capture rates necessitate numerical inversion of the Laplace transform expression. Some exact results are presented for delocalized initial distributions of the particle in the case of perfect absorption. To study arbitrary capture rate we employ a numerical inversion of the Laplace transform 
and verify the calculations through a direct numerical procedure from the partial differential equation consisting of the Smoluchowski equation augmented by the localized capture term. The direct numerical procedure is explained in the Appendix. The advantage of the work reported in Section 3 is the explicit analytic (Whittaker function) expression we provide and well as various analytically obtained time dependences; its disadvantage is that it leads to generally predictable observations with little surprise.

By contrast, Section 4 uncovers interesting phenomena that depend on the relative location of the trap and the attractive center. There, on the basis of a numerical study, we provide analysis for several different locations of the  trap and particle and discover an effect which is perhaps non-intuitive. In Section 5 we present two additional approaches to the problem. One is through consideration of a \emph{transfer rate} we specially define for the purpose at hand, following developments in exciton dynamics theory~\cite{KENKRE:1983}. The other, which presupposes as an approximation that the parameters of the system are such that the equilibrium Smoluchowski distribution is attained quickly, describes the capture efficiency as dependent on that equilibrium distribution. Several new insights into the physics of reaction-diffusion phenomena in the presence of a potential become clear in Sections 4 and 5. In Section 6 we present concluding remarks.

Our point of departure in the \emph{absence of the capture term} is a generalization of the diffusion equation, viz., the Smoluchowski equation, that has the form 
\begin{equation}
\frac{\partial P(x,t)}{\partial t}=\frac{\partial }{\partial x}\left(\gamma x P(x,t)+D\frac{\partial P(x,t)}{\partial x}\right),
\label{Smoluchowski_notrap}
\end{equation}
where $P(x,t)$ is the probability to find the particle at position $x$ and time $t$. The diffusion constant is $D$, the rate at which the particle tends to return to the potential center (taken to be the origin without loss of generality) is $\gamma$.

The propagator of this equation, i.e., the solution for $P(x,0)=\delta (x-x_0)$, to be denoted by the symbol $\Pi(x,x_0,t)$, can be obtained by Fourier transforming the equation and solving the resulting first order partial differential equation by the method of characteristics. See, e.g., ~\cite{RISKEN:1989:ID199,Reichl:2009}:
\begin{equation}
\Pi(x,x_0,t)=\frac{e^{-\frac{\left(x-x_0e^{-\gamma t}\right)^2}{4D\mathcal{T}(t)}}}{\sqrt{4\pi D\mathcal{T}(t)}},
\label{prop}
\end{equation}
where $\mathcal{T}$ is a function of time given by
\begin{equation}
\mathcal{T}(t)=\frac{1-e^{-2\gamma t}}{2\gamma}.
\label{fancyT}
\end{equation}
The solution shows transparently that, wherever it is initially placed, the particle tends to move to the origin at rate $\gamma$ but, as a result of the diffusion that it also undergoes, ends up, in the steady state, occupying a Gaussian  of width proportional to the square root of the ratio of the diffusion constant to $\gamma$. 

\section{Survival Probability}
We are generally interested in the particle survival probability 
\begin{equation}
Q(t)=\int_{-\infty}^{+\infty}dx\,P(x,t)
\end{equation}
when a term  involving the capture parameter $\mathcal{C}_1$ is appended to the Smoluchowski equation to represent capture of the particle at the point $x=x_r$:
\begin{equation}
\begin{split}
\frac{\partial P(x,t)}{\partial t}=&\frac{\partial }{\partial x}\left(\gamma x P(x,t)+D\frac{\partial P(x,t)}{\partial x}\right)\\&-\mathcal{C}_1\delta(x-x_r)P(x,t).
\end{split}
\label{Smoluchowski_trap}
\end{equation}
The use of the standard defect technique~\cite{KENKRE:1983,KENKRE:1980:ID176,KENKRE:1980:ID280, KENKRE:1982:ID259, KENKRE:1983:ID198, KENKRE:1985}  yields the probability density in the Laplace domain (tildes denote the Laplace transform and $\epsilon$ is the Laplace variable),
\begin{equation}
\begin{split}
\widetilde{P}(x,\epsilon)=&\int_{-\infty}^{+\infty}dx_0 \widetilde{\Pi}(x,x_0,\epsilon)P(x_0,0)
\\&-\widetilde{\mathcal{M}}(x_r,\epsilon)\widetilde{\Pi}(x,x_r,\epsilon)
\\& \times\int_{-\infty}^{+\infty}dx_0\widetilde{\Pi}(x_r,x_0,\epsilon)P(x_0,0)
\end{split}
\end{equation}
where
\begin{equation}
\widetilde{\mathcal{M}}(x_r,\epsilon)=\frac{1}{(1/\mathcal{C}_1)+\widetilde{\Pi}(x_r,x_r,\epsilon)}.
\end{equation}
Integrating this expression over all space, one obtains the Laplace-domain result for the survival probability for arbitrary initial conditions:
\begin{equation}
\widetilde{Q}(\epsilon)=\frac{1}{\epsilon}\left[1-\left(\frac{\int_{-\infty}^{+\infty}dx_0 \,\widetilde{\Pi}(x_r,x_0,\epsilon)P(x_0,0)}{(1/\mathcal{C}_1)+\widetilde{\Pi}(x_r,x_r,\epsilon)}\right)\right].
\label{survival}
\end{equation}
The derivation of Eq. (\ref{survival}) from Eq. (\ref{Smoluchowski_trap}) has appeared multiple times in the literature under multiple authorship. Some of its essential steps  may be found collected in a recent article by two of the present authors \cite{KSinJPC}. Note that the numerator in the parentheses, $\int_{-\infty}^{+\infty}dx_0 \,\widetilde{\Pi}(x_r,x_0,\epsilon)P(x_0,0),$ is the homogeneous solution (solution in the absence of capture) at the trap site $x_r$ with the given initial particle placement  $P(x_0,0)$, and that, if the absorption by the trap is perfect (infinite capture rate), the term $1/\mathcal{C}_1$ is identically zero. 

If the initial condition of the particle placement is localized at a single point $x_0$, Eq. (\ref{survival}) reduces to
\begin{equation}
\widetilde{Q}(\epsilon)=\frac{1}{\epsilon}\left[1-\left(\frac{\widetilde{\Pi}(x_r,x_0)}{(1/\mathcal{C}_1)+\widetilde{\Pi}(x_r,x_r)}\right)\right],
\label{locsurvival}
\end{equation}
where (and henceforth) we drop the specification of $\epsilon$ explicitly in the arguments in the right hand side.  The starting point for the calculations we present below is the conjunction of Eq. (\ref{locsurvival}) with Eq. (\ref{prop}).

\section{Centrally Placed Trap: Analytic Solution}
The complexity of the Smoluchowski propagator Eq. (\ref{prop}) makes it difficult or impossible to obtain analytic expressions in most cases. If, however, the trap is located at the attractive center of the potential, $x_r=0,$ progress can be made because the Laplace transforms of the propagators appearing in $\widetilde{Q}(\epsilon)$ can be computed explicitly in terms of the Whittaker function. The survival probability for this case is, in the Laplace domain, 
\begin{equation}
\widetilde{Q}(\epsilon)=\frac{1}{\epsilon}\left[1-\left(\frac{\widetilde{\Pi}(0,x_0)}{(1/\mathcal{C}_1)+\widetilde{\Pi}(0,0)}\right)\right].
\label{trapcenter}
\end{equation}
Putting $x_r=0$ in the general expression for the Laplace transform of the Smoluchowski propagator Eq. (\ref{prop}),
$$\int_0^{\infty}dt\,\left[\frac{e^{-\frac{\left(x_r-x_0e^{-\gamma t}\right)^2}{4D\mathcal{T}(t)}}}{\sqrt{4\pi D\mathcal{T}(t)}}\right]e^{-\epsilon t},$$
one finds, from a table of Laplace transforms~\cite{ROBERTS:1966:ID189} or otherwise, that 
\begin{equation}
\begin{split}
\widetilde{\Pi}(0,x_0)=\frac{(\gamma\tau_1)^{-1/4}}{2\sqrt{\pi}\;\sigma\gamma} e^{(\gamma\tau_1/2)} \Gamma \left( {\frac{\epsilon }{{2\gamma }}} \right)W_{\frac{1}{4} - \frac{\epsilon }{{2\gamma }},\frac{1}{4}}(\gamma\tau_1).
\end{split}
\label{LTpi0x0}
\end{equation}
Here, $\tau_1=x_0^2/2D$ is the time \cite{taunote} the particle would take to move via pure diffusion from its initial location to the attractive center; the dimensionless quantity  $\gamma\tau_1$ is therefore the ratio of that diffusion time to $1/\gamma$, the time characteristic of motion resulting purely from the pull of the potential.  On defining $\sigma=\sqrt{2D/\gamma}$ which is the width of the equilibrium distribution of the trap-less Smoluchowski equation \cite{width}, we see that $\gamma\tau_1$ can be given another physical interpretation: it is identical to the square of the initial location of the particle to the Smoluchowski equilibrium width:
\begin{equation}
\gamma\tau_1=\left(\frac{x_0}{\sigma}\right)^2.
\label{gammat}
\end{equation}
The $W$ in Eq. (\ref{LTpi0x0}) is the Whittaker W-function defined in Ref.~\cite{Abramowitz:1970} as 
\[
W_{\kappa,\mu}(z)=e^{-\frac{z}{2}}z^{\frac{1}{2}+\mu}U\left(\frac{1}{2}+\mu-\kappa,1+2\mu,z\right), |arg z|<\pi 
\]
in terms of the confluent hypergeometric function
\[
U\left(a,b,c\right)=\frac{1}{\Gamma(a)}\int\limits_0^{\infty} { e^{-ct}t^{a-1}\left(1+t\right)^{b-a-1}dt}.
\]
The transform of the other propagator in Eq. (\ref{trapcenter}) is even easier to calculate. One puts $x_0=0$ in the above expression, or directly computes the integral 
$$\frac{1}{\sigma\sqrt{\pi}}\int_0^{\infty}dt\,\frac{e^{-\epsilon t}}{\sqrt{1-e^{-2\gamma t}}},$$
to get
\begin{equation}
\widetilde{\Pi}(0,0)=\frac{1}{\epsilon \sigma}\frac{\Gamma\left(\frac{\epsilon}{2\gamma}+1\right)}{\Gamma\left(\frac{\epsilon}{2\gamma}+\frac{1}{2}\right)}\\= \frac{1}{2\gamma \sigma\sqrt{\pi}}B\left(\frac{\epsilon}{2\gamma},\frac{1}{2}\right),
\label{LTpi00}
\end{equation}
where $\Gamma(n)$ is the Gamma function and $B(z,w)=\Gamma(z)\Gamma(w)/\Gamma(z+w)$ is the Beta function. 

Substitution of Eqs. (\ref{LTpi0x0}) and (\ref{LTpi00}) into the prescription given in Eq.(\ref{trapcenter}) provides an exact expression for the total survival probability in the Laplace domain:
\begin{equation}
\widetilde Q(\epsilon)=\frac{1}{\epsilon}\left[1-\frac{ (\gamma\tau_1)^{ -1/4} e^{(\gamma\tau_1/2)} \Gamma \left( {\frac{\epsilon }{{2\gamma }}} \right)W_{\frac{1}{4} - \frac{\epsilon }{{2\gamma }},\frac{1}{4}}(\gamma\tau_1)} { \xi + B\left(\frac{\epsilon}{2\gamma},\frac{1}{2}\right)}\right].
\label{centertrapW}
\end{equation}
The denominator of the second term in the square brackets is a sum of a dimensionless motion quantity that appears in the form of the Beta function, and a dimensionless capture parameter
\begin{equation}
\xi=2\sqrt{\pi}\left(\frac{\gamma\sigma}{\mathcal{C}_1}\right)=2\sqrt{2\pi}\left(\frac{\sqrt{\gamma D}}{\mathcal{C}_1}\right),
\label{captureparameter}
\end{equation}
which is inversely proportional to $\mathcal{C}_1$. The dimensionless ratio in the parentheses in Eq. (\ref{captureparameter}) compares a time for capture to a time for motion and represents the extent of \emph{imperfectness} of absorption. If $\xi$ vanishes, that imperfection vanishes and one has a perfect absorber; if $\xi$ is large, one has weak capture.

Equation~(\ref{centertrapW}) is one of the primary results of our paper. While generally the Laplace-inversion of its right hand side cannot be done analytically, and necessitates numerical procedures, for perfect absorption a surprising reduction occurs. 

\subsection{Analytic Inversion for Perfect Absorption}
\begin{figure}[h] 
  \centering
  \includegraphics[width=0.5\textwidth]{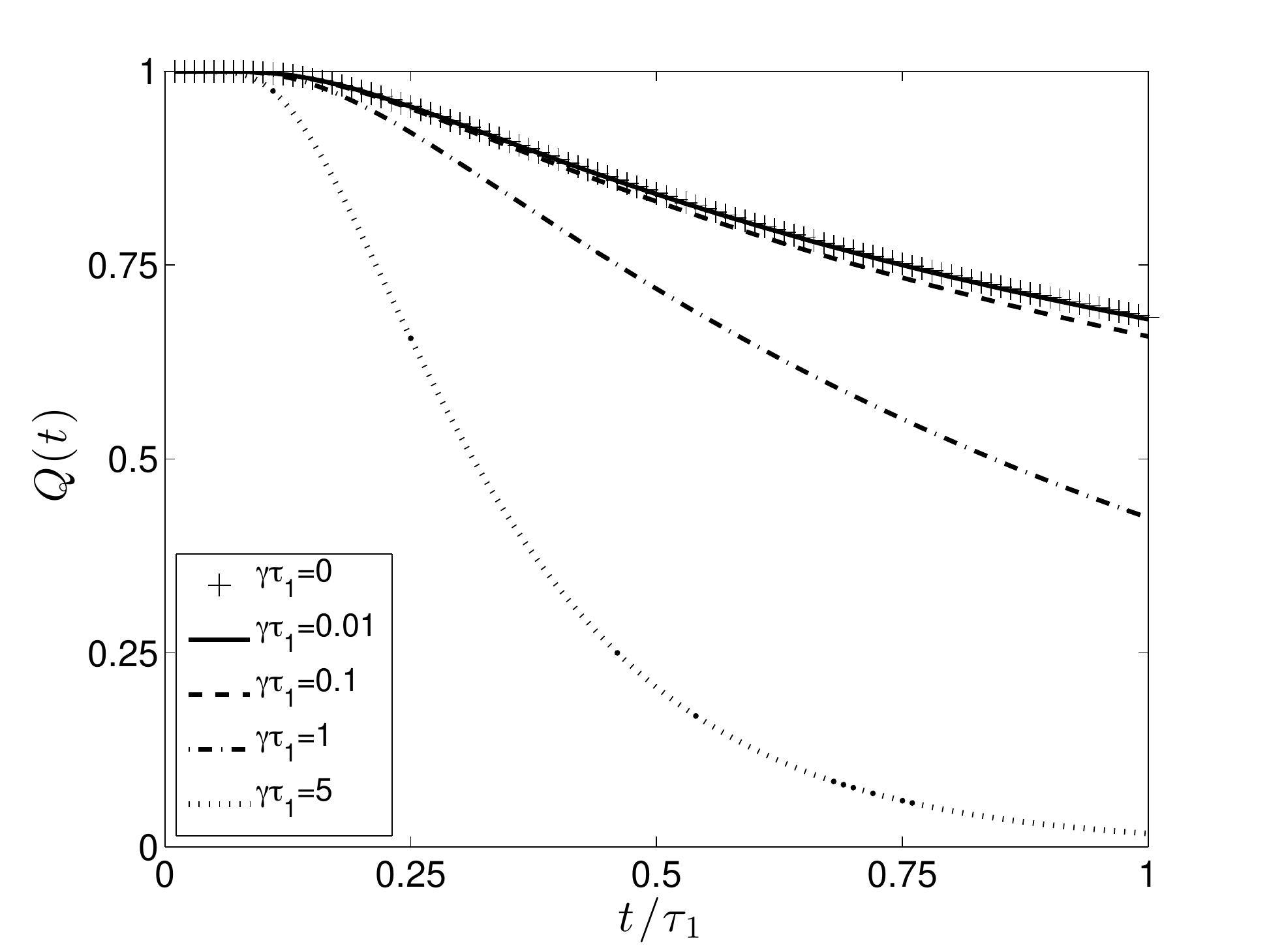}
  \caption{Survival probability for perfect capture at the trap located at center. The curves converge to the diffusive limit, Eq. (\ref{tomdickharry}), (topmost curve) as $\gamma \rightarrow 0$. $Q(t)$ is shown for several $\gamma\tau_1$ values: 5, 1, 0.1, 0.01, 0.001.}
  \label{fig:Figure1}
\end{figure}

For perfect absorption, $\mathcal{C}_1 \rightarrow \infty$ making $\xi=0$ vanish, Eq. (\ref{centertrapW}) reduces to
\begin{equation}
\widetilde Q(\epsilon)=\frac{1}{\epsilon}-{\frac{\Gamma\left(\frac{\epsilon+2\gamma}{2\gamma}\right)\Gamma\left(\frac{\epsilon}{2\gamma}\right)}{2\gamma\sqrt{\pi}\;(\gamma\tau_1)^{1/4}\;\Gamma\left(\frac{\epsilon+\gamma}{2\gamma}\right)}e^{\frac{\gamma\tau_1}{2}}W_{-\frac{1}{4},\frac{1}{4}}\left(\gamma\tau_1\right)},
\end{equation}
and inversion back to the time domain is readily possible:
\begin{equation}
Q(t) = 1 - \left( \frac{{{\mathop{\rm e}\nolimits} ^{2\gamma t}  - 1} }{\pi^2\gamma\tau_1}\right)^{\frac{1}{4}} e^{\frac{{ - \gamma\tau_1}}{{{e^{2\gamma t} - 1}}}}~W_{- \frac{1}{4},\frac{1}{4}}\left(\frac{\gamma\tau_1}{{e^{2\gamma t} - 1}}\right).
\label{chap6_quadratic_Qw_time}
\end{equation}
Here we use well-known scaling and shift rules along with the result~\cite{ROBERTS:1966:ID189} that the Laplace transform of
$$e^{ - \frac{a}{2}} \left( {1 - {\mathop{\rm e}\nolimits} ^{ - t} } \right)^{ - \mu } e^{\frac{{ - a/2}}{{\left(e^t - 1\right)}}} W_{\mu ,\nu }\left(\frac{a}{{e^t - 1}} \right)$$
is
$$\frac{{\Gamma \left( {\epsilon  + 1/2 + \nu } \right)\Gamma \left( {\epsilon  + 1/2 - \nu } \right)}}{{\Gamma \left( {\epsilon ' + 1 - \mu } \right)}}W_{-\epsilon ,\nu }\left(a\right).$$
We see that the survival probability in the time domain Eq. (\ref{chap6_quadratic_Qw_time}) also involves the Whittaker function with an argument that is itself a function of time. Furthermore,  $W(t)$, in the form it appears in Eq.(\ref{chap6_quadratic_Qw_time}), can be defined in terms of the complementary error function~\cite{Abramowitz:1970},
\begin{equation}
W_{- \frac{1}{4},\frac{1}{4}}(z)=\sqrt{\pi}z^{1/4}e^{z/2}\text{erfc}(\sqrt{z}).
\end{equation}
This has the remarkable consequence that, for perfect absorption, we can derive the simple result
\begin{equation}
Q(t) = \text{erf}\left( \frac{x_0/\sigma}{\sqrt{e^{2\gamma t}-1}}\right)=\text{erf}\sqrt{ \frac{\gamma \tau_1}{e^{2\gamma \tau_1 (t/\tau_1)}-1}}.
\label{centertrapperfect}
\end{equation}
There is much that can be said about Eq. (\ref{centertrapperfect}). The error function behavior ensures that the survival probability does not change much initially but only after a threshold time has elapsed. All time derivatives of Q(t) of finite order vanish at the origin. The threshold time might be taken to signify that the particle has arrived at the trap. After that, the time scale for the evolution of $Q(t)$ is  generally $1/\gamma$ but, in the limit that this time becomes infinite (infinitely flat potential, $\gamma \rightarrow 0$), the characteristic time becomes $\tau_1$. One sees here transparently the transition from potential-induced motion to the trap to diffusive motion. In that diffusive limit (no potential), Eq. (\ref{centertrapperfect}) reduces to the well-known result~\cite{CARSLAW:1959:ID186,SPOUGE:1988:ID187,REDNER:1990:ID54,REDNER:2001:ID188,BLYTHE:2003:ID227, KSinJPC}
\begin{equation}
Q(t) = \text{erf}\left( \sqrt {\frac{\tau_1 }{2t}} \right).
\label{tomdickharry}
\end{equation}
Figure \ref{fig:Figure1} shows the time dependence of the survival probability for perfect absorption (infinite $\mathcal{C}_1$) for five values (5, 1, 0.1, 0.01, 0.001) of $\gamma\tau_1$. The curves converge to a limit (curves for the lowest two values of $\gamma\tau_1$ practically coincide) that represents pure diffusive motion with no potential pull. 

\subsection{Delocalized Initial Particle Distribution: Superposition}
Once the exact solution is known for a point initial condition $P(x_0)=\delta(x-x_0)$ for particle placement, one can solve the problem for any initial condition by summing the results. Thus, provided one has the perfect absorption case, the principle of superposition yields
\begin{equation}
Q(t)=\int_{-\infty}^{\infty}dx_0P(x_0) \text{erf}\left( \frac{x_0/\sigma}{\sqrt{e^{2\gamma t}-1}}\right).
\label{transform}
\end{equation}
The survival probability in Eq. (\ref{transform}) for this central-trap perfect-absorber system may be viewed  as a transform of the initial probability distribution $P(x_0)$ of the particle. In a number of useful situations the initial particle distribution is non vanishing only on one side of the potential center. Then the lower limit in the integration of Eq. (\ref{transform}) becomes $0$ and a situation akin to the Laplace transform occurs. The error function  takes the place of the exponential in the Laplace transform, and the quantity $(\sigma\sqrt{e^{2\gamma t}-1})^{-1}$ plays the role of the transform variable $\epsilon$. 

We display two useful consequences of this transform. For an initial exponential distribution $P(x_0)=(1/d)\exp(-x_0/d)$ only on one side, i.e., for $x_0 > 0$ (and vanishing $P(x_0)$ elsewhere), with characteristic distance $d$, the survival probability is
\begin{equation}
Q(t)=e^{\zeta^2(t)} \text{erfc}\left(\zeta(t) \right)
\label{center_random}
\end{equation}
where $\zeta(t)=(\sigma/2d)\sqrt{e^{2\gamma t}-1}$. For an initial Rayleigh distribution $P(x_0)=(x_0/d^2)exp[-x_0^2/(2d^2)]$ for $x_0 > 0$ (and vanishing $P(x_0)$ elsewhere), we get
\begin{equation}
Q(t)=\left[1+\left(\sigma^2/2d^2\right)(e^{2\gamma t}-1)\right]^{-1/2}.
\label{center_Rayleigh}
\end{equation}
The first of the distributions, often called the \emph{random} or Poisson distribution, arises often and can describe, for instance, the initial placement of coalescing signaling receptor clusters in immune mast cells~\cite{SPENDIER:2010:ID143}.  The second distribution is a biased Poisson distribution which also occurs in several physical systems. We have mentioned both of them because the first concentrates the initial placement of the particle near the attractive center while the second shifts it away by a finite amount. We have used $d$ to denote the average value $\int x_0P(x_0)dx_0$ in both cases. 

\begin{figure}[h] 
  \centering
  \includegraphics[width=0.5\textwidth]{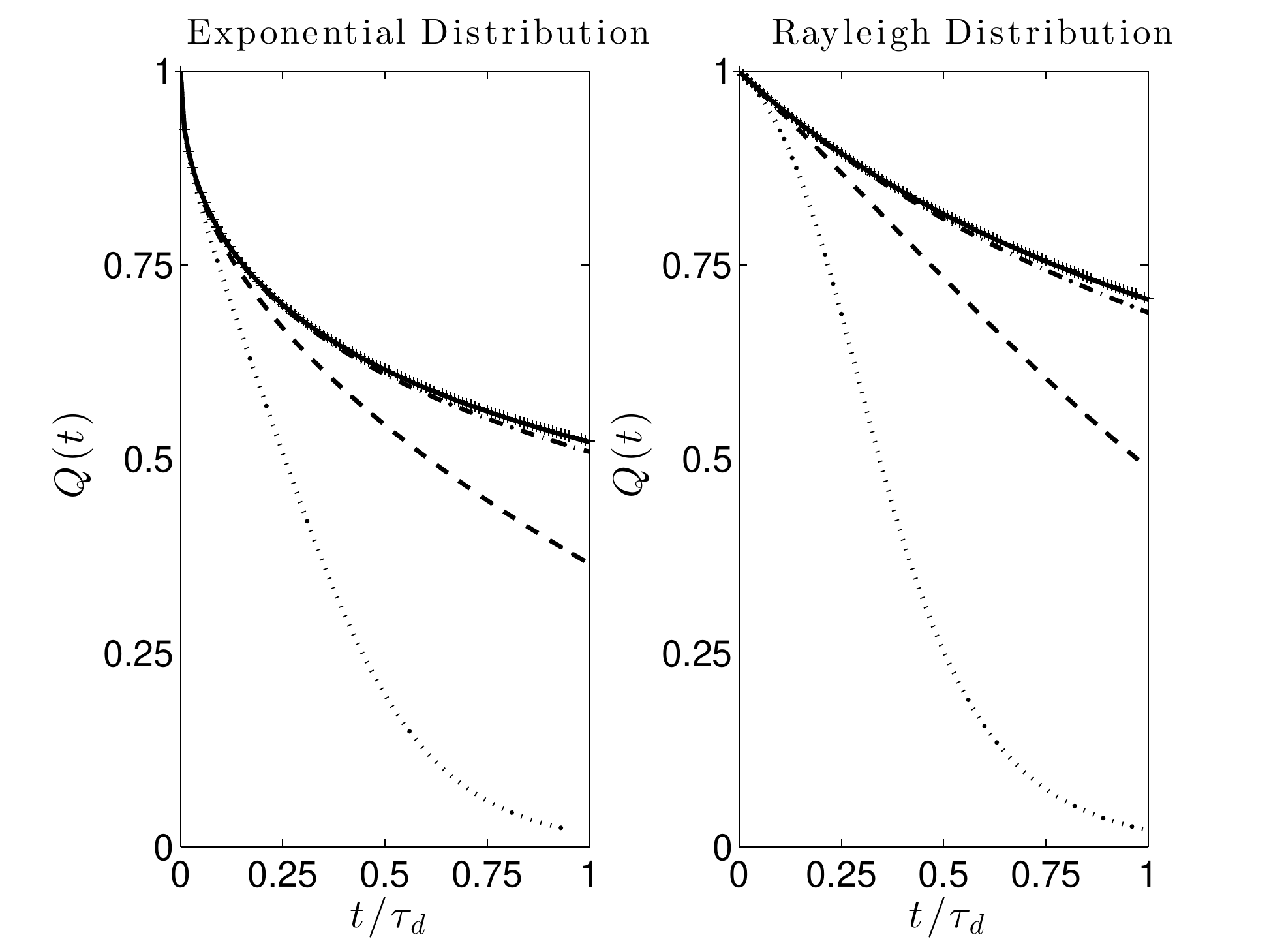}
  \caption{Survival probability for exponential (left panel) and Rayleigh (right panel) initial particle distributions as given in Eqs. (\ref{center_random}), (\ref{center_Rayleigh}). In each panel, the curves correspond to $(d/\sigma)^{2}$ = 5, 1, 0.1, 0.01, and 0.001, from the bottom to the top curve, respectively. Here $\tau_d=d^2/2D$ and $d$ is the characteristic distance of the initial distribution. The near-origin behavior of Q(t) is substantially different from that in Fig. \ref{fig:Figure1}. See text.}
  \label{fig:Figure2}
\end{figure}

Figure \ref{fig:Figure2} shows the two cases of the survival probability for the two initial particle distributions. In both of them we see that the $Q(t)$ curves converge to the pure diffusive limit (top line). The characteristic time $\tau_d$ in the units of which $t$  is plotted in these curves equals $d^2/2D$, i.e., is the time the particle would take to traverse diffusively the characteristic distance $d$ for each of the distributions.

A noteworthy feature of Eqs. (\ref{center_random}) and (\ref{center_Rayleigh}), and of Fig. \ref{fig:Figure2}, is the loss of the  reverse Arrhenius behavior near the origin  (derivatives of all finite orders vanishing at the origin) brought about by superposition of contributions from multiple initial locations of the particles: Q(t) curves, while \emph{totally flat} as $t \rightarrow 0$ in Fig. \ref{fig:Figure1}, change through superposition to non-drastic variation near the origin in Fig. \ref{fig:Figure2}. The mathematical mechanism for this conversion is the removal of the isolated essential singularity by integration and is essentially the one encountered in the temperature ($T$) dependence of the specific heat of insulators. It is well-known \cite{debye} that, through a superposition of activated Einstein contributions, each of which fails to describe the correct near-origin temperature behavior, the Debye theory succeeds in predicting the correct, dimension-driven $T^3$ dependence. In our present problem, the near-origin \emph{time-}dependence of $Q(t)$ plays the role of the near-origin \emph{temperature-}dependence of the specific heat.

Equation~(\ref{centertrapW}) for arbitrary capture rate,  the demonstration of the analytic reduction to the perfect absorber result Eq. (\ref{centertrapperfect}), and the superposition results Eqs. (\ref{transform}), (\ref{center_random}), and (\ref{center_Rayleigh}) are  our main results in this subsection. We have found that a passing mention of the perfect-absorber central-trap localized result for the centrally placed trap, Eq. (\ref{centertrapperfect}), has appeared in a previous analysis~\cite{BAGCHI:1983:ID275} on the effect of viscosity on electronic relaxation in solution.

\subsection{Numerical Inversion for Noninfinite Capture}
Numerical Laplace inversion of Eq. (\ref{centertrapW}) becomes necessary for finite $\mathcal{C}_1$, equivalently for non-vanishing $\xi$. We use standard inversion routines~\cite{Gaver,Abate} and get satisfactory coincidence with the direct numerical solution (via discretization) of the partial differential equation except when the latter is inaccurate because the Smoluchowski width is too narrow. See discussion in the Appendix. Because of the confidence gained thus in the inversion procedure, we use it for numerical calculations throughout the rest of the paper.

We have explored the survival probability for various values of the capture rate $\mathcal{C}_1$, equivalently of the dimensionless parameter $\xi$. We have uncovered no surprises. A stronger capture rate makes $Q(t)$ decrease faster as expected. We have not found it instructive to display the resultant figures. We emphasize, however, that our procedure can produce the evolution of the survival probability for arbitrary capture. 

\section{Analysis for arbitrary locations}
Situations in which the attractive center, the trap, and the initial placement of the moving particle are at \emph{arbitrary}  locations with respect to one another, are rich in their outcome. This is expected. For instance, one might argue that, if the initial location of the particle lies in between the potential center and the trap, the pull provided by the confining potential would tend to act counter to the phenomenon of trapping and that the potential would thus hinder trapping and enhance survival. Yet, since at equilibrium, the particle in the trap-less situation would tend to occupy an extent around the potential center given by the Smoluchowski width, one might expect survival, when the trap is present, to depend on whether the distance of the trap from the potential center is disparate with respect to the Smoluchowski width. Which effect takes over in a given set of circumstances? These interesting situations are difficult, or even generally impossible, to study via analytic 
solutions. To investigate them, the straightforward way is to use a numerical program that starts with Eq. (\ref{survival}) or Eq. (\ref{locsurvival}), depending on the initial condition (general or localized), to substitute in it the Laplace transforms of the propagators evaluated numerically from Eqs. (\ref{prop}) and (\ref{fancyT}), and to perform the numerical Laplace inversion by standard methods  to produce the final time-dependent survival probability. We pursue this program systematically in this Section. 

\subsection{Symmetrical Placement of Trap and Particle}
\begin{figure}[h] 
   \centering
   \includegraphics[width=1\columnwidth]{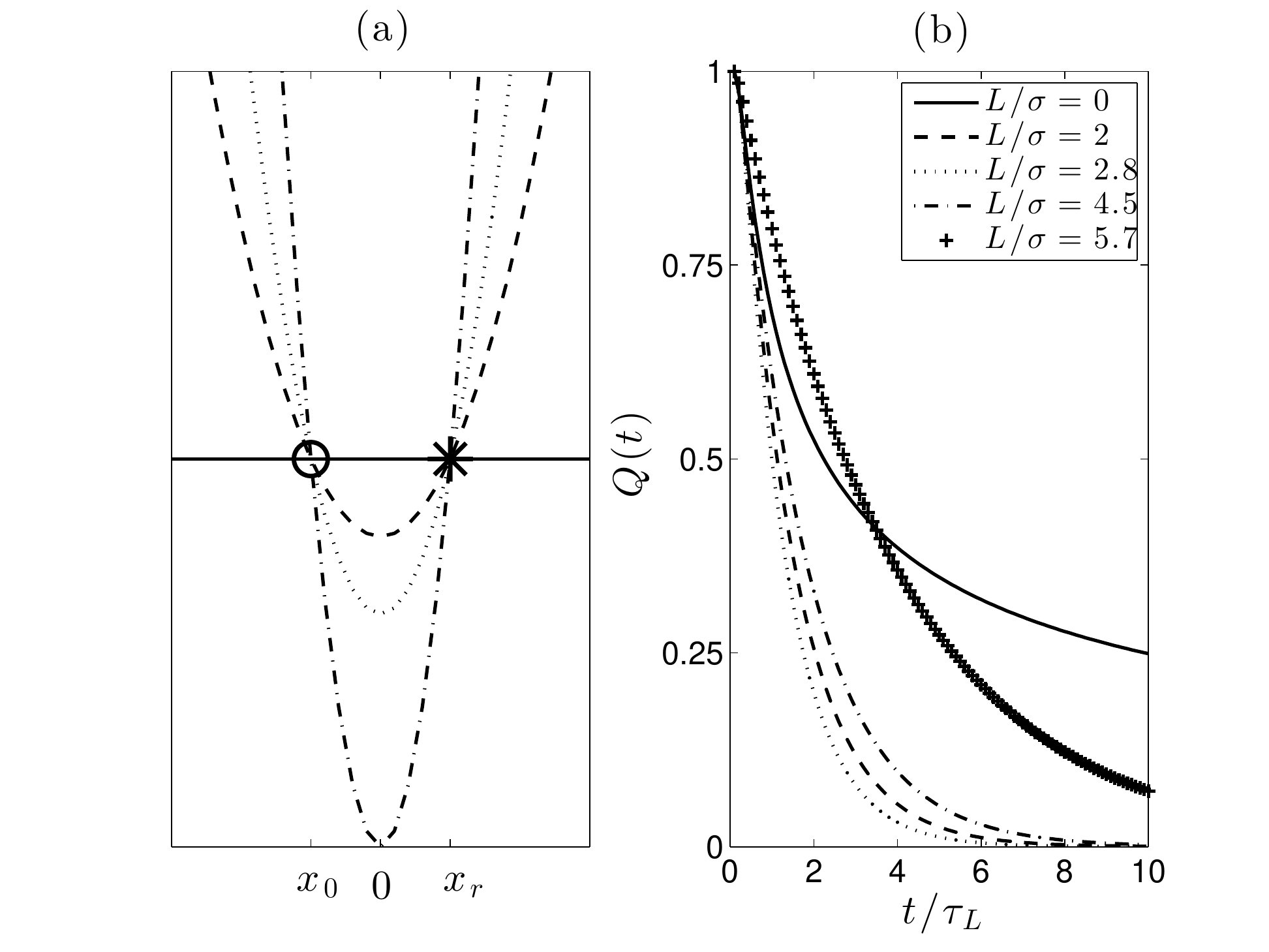}
   \caption{Non-monotonic dependence of capture efficiency on potential steepness for symmetrical placement of trap and particle. Left panel represents the situation visually. Right panel shows the non-monotonic effect as the decay of $Q(t)$ is enhanced by increasing the potential steepness but then hindered on further increase. Curves are labeled by $L/\sigma$, the ratio of the distance between trap and initial location of particle to the Smoluchowski width. Four of the traces in the right panel (for $L/\sigma=0, 2, 2.5$ and $4.5$) correspond respectively to the potential curves in the left panel.}
   \label{fig:Figure3}
 \end{figure}

Let us first consider the case of no potential ($\gamma=0$),  the trap placed at $x_r=L/2$ and the initial location of the particle at $x_0=-L/2$ so that the distance between the two is $L$. The survival probability $Q(t)$ is given by the well-known expression (\ref{tomdickharry}) valid for a diffusion rather than a Smoluchowski equation, with $\tau_1$ replaced by $\tau_L=L^2/2D$. Let us now introduce a potential with its attractive center precisely midway between the trap and the particle (the potential center is at $0$), see Fig. \ref{fig:Figure3}a, and examine the time dependence of the survival probability as the potential steepness measured by $\gamma$, or more conveniently  the dimensionless ratio $L/\sigma$, is varied. We display the results in Fig. \ref{fig:Figure3}b. 

Starting with the pure diffusive case $L/\sigma=0$, for which $\sigma$ is infinite and the survival probability is given by Eq. (\ref{tomdickharry}), we see that increase of potential steepness, equivalently of $L/\sigma$, has a remarkable non-monotonic effect. Small increase makes capture more efficient but beyond a certain value it has the opposite effect. Why does this happen? The presence of a potential surely makes the particle move faster, at least initially, towards the trap as it travels to the attractive center. However, past the attractive center, the motion towards the trap is uphill and therefore \emph{hindered} by the potential. The introduction of the potential thus has both a favorable and an unfavorable effect on capture. 

There is an approximate but instructive way to think about what is happening by comparing where the trap lies in relation to the Smoluchowski width. The potential pull tends to bring the probability density at the trap location to its equilibrium value in the absence of the capture. The dependence of this value (indeed the value at any location which is not the potential center) on the steepness of the potential is non-monotonic. This will become quantitatively clear in Section 5, Eq. (\ref{approxdqdt}), and Fig. \ref{fig:Figure6} below.

\subsection{One-sided Placement of Trap and Particle}

\begin{figure}[h] 
 \begin{center}
  \includegraphics[width=1\columnwidth]{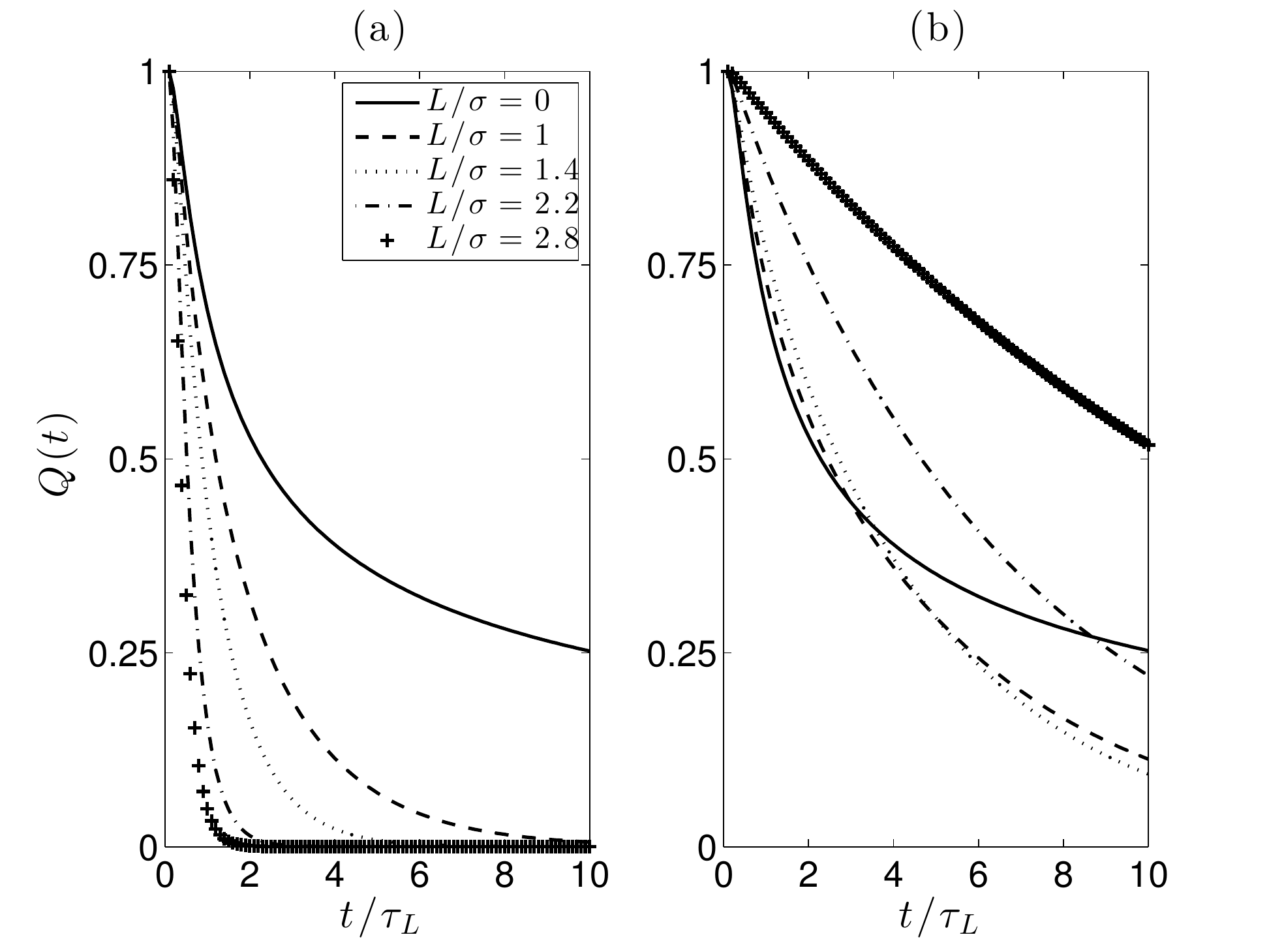}  
  \caption{Difference in the behavior of $Q(t)$ for uphill and downhill placement of trap with respect to initial particle location. The trap is at the potential center in the left panel but uphill in the right panel (initial particle location being at the potential center in this case). The effect of increasing the potential steepness is monotonic in the former but displays interesting features as in Fig. \ref{fig:Figure3} in the latter.}
  \label{fig:Figure4}
 \end{center}
\end{figure}

Let us now place the trap at the center and the particle placed initially uphill at some distance $L$. This means $x_r=0$ and $x_0=L$. Nothing particularly interesting emerges as a potential is introduced (with attractive center at the origin, as earlier) and its steepness is varied: steeper potentials make trapping easier. On the other hand, if we reverse the positions of the trap and the particle, so that $x_r=L$ and $x_0=0$, interesting non-monotonic behavior is met with again with variation in potential steepness. These two cases of uphill particle and uphill trap (respectively) are shown in Fig. \ref{fig:Figure4}, $\tau_L$ being, as in Fig. \ref{fig:Figure3}, the time taken by the particle to traverse the distance from its initial placement to the trap under purely diffusive condition.

Why is there no symmetry in the effects of the placement of the stationary trap and the moving particle? After all, survival probability depends merely on their meeting. The answer is obvious. A particle placed uphill with the trap at the potential center is always helped by the potential steepness to get faster to the trap. The curves in the left panel of Fig.~\ref{fig:Figure4} therefore show more capture as the potential steepness increases.

Our investigations have uncovered further subtle effects of varying the relative locations of the trap, particle and potential center. We intend to present their analysis elsewhere.

\section{Additional Methods of Analysis and Intelligent Design}
We discuss two additional approaches for the analysis of the problem under consideration. Coupled with the numerical and exact investigations displayed in the previous sections, these two help in understanding the phenomena that occur in this field of reaction diffusion and assist in the intelligent design of devices based on them.

\subsection{Transfer Rate Approach}
Laplace inversion programs sometimes have the reputation of being inaccurate if the corresponding time-dependent function possesses oscillations. The cases we have examined do not show this feature. Nevertheless, it is helpful to develop a method to simplify the analysis and perform calculations without (numerical) inversion of Laplace transforms. Such a method has been previously applied by one of the present authors in the study of exciton transport in molecular crystals~\cite{KENKRE:1980:ID176}  and proceeds through the definition of a transfer rate $k$ from the host system in which the particle moves to the trap. We describe it below.

If the survival probability were an exponential in time, i.e., $Q(t)=\exp(-kt),$  the transfer rate would be given precisely by $k=1/\int_0^{\infty}Q(t)dt$. For many systems the integral in the denominator blows up. In such cases, it is helpful to introduce a lifetime $\tau$ into our starting Eq. (\ref{Smoluchowski_trap}) so that the probability density, which we will now call $p(x,t),$ obeys
\begin{equation}
\begin{split}
\frac{\partial p(x,t)}{\partial t}+\frac{p(x,t)}{\tau}=&\frac{\partial }{\partial x}\left(\gamma x p(x,t)+D\frac{\partial p(x,t)}{\partial x}\right)\\&-\mathcal{C}_1\delta(x-x_r)p(x,t).
\end{split}
\label{smoludecay}
\end{equation}
If the moving particle is an excitation (e.g., Frenkel exciton as in the case of photosynthesis) the presence of a finite lifetime can be a consequence of the physics of the system: it corresponds often to the radiative, and sometimes to nonradiative, decay of the particle as the excitation turns into a photon or disappears in other ways. If the moving particle is an animal in the context of ecology, $\tau$ could be its actual lifetime, its finiteness caused by predators or natural causes. If the physics of the system does not include a finite lifetime for the particle, $\tau$ in our present analysis should be regarded as simply a probe time.

Now, given that the solution $p(x,t)$ of Eq. (\ref{smoludecay}) is trivially related to the solution $P(x,t)$ of Eq. (\ref{Smoluchowski_trap}) through $p(x,t)=P(x,t)\exp(-t/\tau$), we see, as shown elsewhere \cite{KENKRE:1980:ID176}, that a useful measure of the transfer rate is
\begin{equation}
k=\left[\frac{1}{\widetilde{Q}(\epsilon)}-\epsilon \right]_{\epsilon=1/\tau}.
\end{equation}
It is generally the value, at $\epsilon=1/\tau,$ of the Laplace transform of the memory $\kappa(t)$ in the expression of the evolution of the survival probability $Q(t)$ written as
\begin{equation}
\frac{dQ(t)}{dt}+\int_0^t dt'\,\kappa(t-t')Q(t')=0.
\end{equation}
For our specific problem involving the Smoluchowski equation it is given, in terms of the propagators of the equation, by
\begin{equation}
k=\frac{1}{\tau}\left[\frac{\widetilde{\Pi}(x_r,x_0)}{(1/\mathcal{C}_1)+\widetilde{\Pi}(x_r,x_r)-\widetilde{\Pi}(x_r,x_0)}\right]_{\epsilon=1/\tau}.
\label{k}
\end{equation}

The advantage of exploring the survival probability with the help of the transfer rate is that, for a given set of parameter values, $k$ is a single quantity rather than a time-dependent curve as is $Q(t)$. When a lifetime $\tau$ is actually present in the system, the quantity $k$ is perfectly suited for study. When there is no actual $\tau,$ the rate $k$ is to be examined for various values of $\tau$ interpreted as a probe time: small values correspond to probing at short times, while large values correspond to accumulated probing at long times. 

\begin{figure}[t]
  \centering
  \includegraphics[bb=45 203 564 576,width=3.3in,height=2.37in,keepaspectratio]{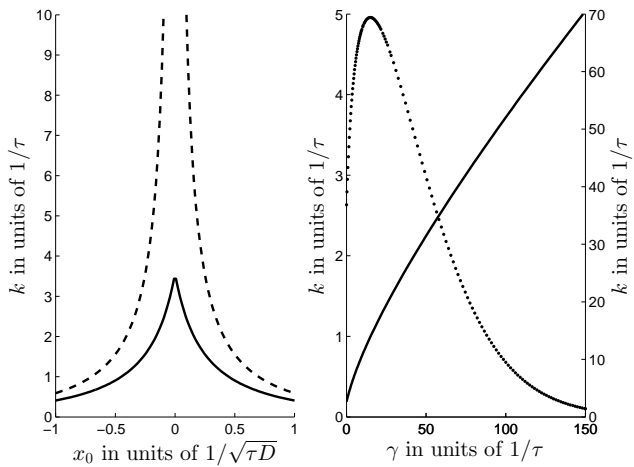}
  \caption{Dependence of the transfer rate $k$. Left panel: The trap is placed at the center of the potential and $k$ is as given in Eq. (\ref{kTrapCenter_imper}). The two curves correspond to two values of the capture parameters $\xi$: $0.1$ (line) and $0$ (dashes). The strength of the attractive potential is fixed to $\gamma\tau = 0.02$. Right panel: The potential steepness represented by $\gamma$ (in units of $1/\tau$) is varied for two configurations of trap and initial position placement showing the non-monotonic effect in one but not the other. Dots (left y-axis) and line (right y-axis) correspond to the respective  placement of the  particle and the trap at the center of the potential, the other being placed uphill. In both cases $\tau_L/\tau = 0.05$ and $\sqrt{D/\tau}/\mathcal{C}_1=0.0032$. Central placement of trap results in monotonic dependence of $k$ on $\gamma$ but central placement of particle leads to the occurrence of an optimum steepness of the potential at which $k$ has a maximum.}
  \label{fig:Figure5}
\end{figure}

Analytic expressions for the transfer rate are possible in the case of a trap at the center of the attractive potential. For arbitrary capture rate, Eq. (\ref{k}) becomes
\begin{equation}
k\tau=\frac{ HW_{\frac{1}{4} - \frac{1}{{2\gamma\tau }},\frac{1}{4}}(\gamma\tau_1)} { \xi + B\left(\frac{1}{2\gamma\tau},\frac{1}{2}\right)-HW_{\frac{1}{4} - \frac{1 }{{2\gamma\tau }},\frac{1}{4}}(\gamma\tau_1)}.
\label{kTrapCenter_imper}
\end{equation}
where
$$H=(\gamma\tau_1)^{ -1/4}e^{(\gamma\tau_1/2)} \Gamma \left( {\frac{1 }{{2\gamma\tau }}} \right).$$
For perfect absorption, $\xi$ vanishes.
The left panel of Fig. \ref{fig:Figure5} depicts the evolution of the transfer rate when the trap is at the center of the potential for two different capture parameters, $\xi=0.1$ (line) and $\xi=0$ (dashes). The transfer rate increases as the initial location of the particle gets closer to the center of the attractive potential and is at a maximum when the particle is at the center, as expected. For perfect absorption it reaches infinity for a particle initially placed at the trap location (dashes) as full capture occurs instantly. Further studies of our expression  show that the width of $k$ as a function of $x_0$ depends  on $\sigma$: as $\sigma$ increases the width decreases. The strength of $k$ depends on the capture rate as well as on the probe lifetime $\tau$. A shorter lifetime or a stronger capture rate both result in an increase in the transfer rate. All effects are as expected.

The right panel of Fig. \ref{fig:Figure5} depicts the dependence of the transfer rate on the potential steepness $\gamma$. We take the initial distance between trap and particle to correspond to $\tau_L/\tau = 0.05$, in one case take $x_r=0$ with $x_0=L$ and in the other, $x_r=L$ with $x_0=0$. When the trap is placed at the center, the transfer rate increases monotonically (line). However, when the trap is placed uphill (dots), a non-monotonic transfer rate emerges as discussed earlier.

\subsection{Approximation for Small Capture Rate}

\begin{figure}[b]
  \centering
  \includegraphics[bb=53 205 556 586,width=3.3in,height=2.5in,keepaspectratio]{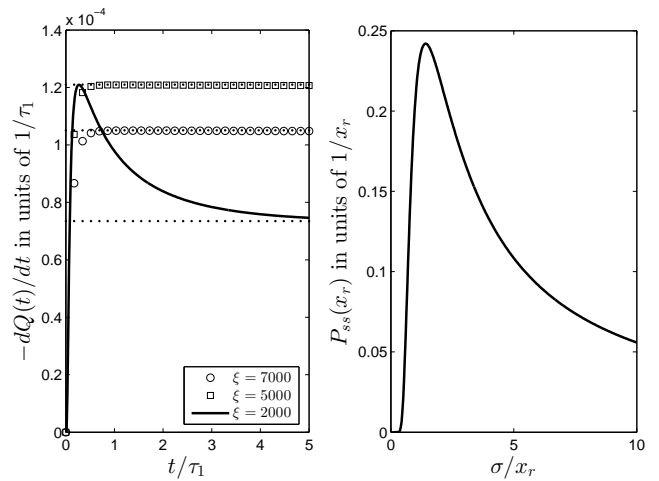}
  \caption{Small capture rate approximation given by Eq. (\ref{approxdqdt}). Left panel:  numerical (exact) solution for the decay rate of $Q(t)$ for three values of $\xi$ as shown. They correspond, respectively, (see text) to $\sigma$ values in the ratio $1:1.4:3.5$. The approximation, which is represented by the three asymptotic (constant) values is seen to describe the evolution adequately at long times. The lowest $\sigma$ value corresponds to circles and the highest to the solid line. Thus, the non-monotonic effect appears at long times. Right panel: $P_{ss}(x_r)$ as a function of $\sigma$ in units of the trap distance from the potential center. Because $P_{ss}(x_r)$  is proportional to the decay rate in the approximation, the fact that it increases, peaks and then decays, is a clear manifestation of the non-monotonicity effect.}
  \label{fig:Figure6}
\end{figure}

The perfect (infinite) capture case has allowed us to present exact, i.e. analytic, solutions in Section 3. The opposite extreme of small capture suggests the following approximation procedure. If $\mathcal{C}_1$ is sufficiently small, $\xi$ in Eq. (\ref{centertrapW}), or more generally $1/\mathcal{C}_1$ in Eq. (\ref{survival}), overwhelms the other term in the respective denominator and leads to a simplification. Working from the more general expression in Eq. (\ref{survival}), we see that, for small capture rate, one may approximate
\begin{equation}
\frac{dQ(t)}{dt}=-\mathcal{C}_1\int_{-\infty}^{\infty}dx_0\,\Pi(x_r,x_0,t)P(x_0,0).
\end{equation}
The time rate of Q(t) is expressed by this approximation to be the negative of the product of $\mathcal{C}_1$ and the probability density at the trap site for the given initial condition but \emph{computed from the homogeneous (i.e., trap-less)} system. Furthermore, under the situation that the relaxation under the potential is sufficiently faster than capture, we may consider (as an approximation) that the steady state distribution $P_{ss}(x_r)$ is achieved  before capture begins, and write
\begin{equation}
\frac{dQ(t)}{dt}=-\mathcal{C}_1P_{ss}(x_r)=-\left(\frac{\mathcal{C}_1}{\sqrt{\pi}}\right)\frac{e^{-(x_r/\sigma)^2}}{\sigma }.
\label{approxdqdt}
\end{equation}
Care must be taken, of course, not to extend this approximation to the point that the survival probability becomes negative. The important point to notice is that Eq. (\ref{approxdqdt}) transparently shows non-monotonicity in the behavior of capture efficiency as a function of the steepness of the potential. Differentiation of $dQ(t)/dt$ with respect to the Smoluchowski width $\sigma$ shows that optimum capture occurs when that width is of the order the trap site distance from the potential center, explicitly when $\sigma=\pm \sqrt{2} x_r$.  This behavior is seen in the plot of the value of the steady state distribution $P_{ss}(x)$ at $x=x_r$ in the right panel of Fig. (\ref{fig:Figure6}). In the left panel, numerically obtained (exact) solutions (circles, squares, solid line) for the decay rate of the survival probability are compared to the approximate predictions of Eq.(\ref{approxdqdt}) (dots). The former are time-dependent as the probability distribution relaxes to the steady state but approach the respective constant values given by the latter at large times. Three cases of the imperfection parameter $\xi$ (see Eq. (\ref{captureparameter})) are shown. Because $\mathcal{C}_1$ and $D$ are held constant for the three cases, these values correspond to values of $\sigma$ that are in the ratio $1:1.4:3.5$ (circles, squares, solid line) respectively. The large time values of the decay rate, displayed as the horizontal asymptotes in the plot, show the non-monotonicity effect.

\section{Concluding Remarks}

The purpose of this paper has been the construction of a reaction diffusion theory, explicitly in 1-d, but  generalizable to higher dimensions, for random walkers moving under an attractive harmonic potential. The appropriate equation is the Smoluchowski equation augmented by capture terms. Our basic starting point is accordingly Eq. (\ref{Smoluchowski_trap}). Our basic result is Eq. (\ref{survival}) and applies to arbitrary strength of capture, not merely to perfect absorption. The amount of departure from perfect absorption is measured by the dimensionless parameter $\xi$ which, as Eq. (\ref{captureparameter}) details, is inversely proportional to the capture rate $\mathcal{C}_1$ and directly proportional to the Smoluchowski width $\sigma$ and to the rate $\gamma$ with the which the particle is pulled to the attractive center. 
Although our theoretical development is thus  capable of addressing arbitrary amounts of capture, perfect absorption yields several exact (analytical) results when the trap location coincides with the attractive center. We have presented them in Section 3. They include, as Figs. \ref{fig:Figure1} and show, the explicit demonstration of the effect of the potential on the survival probability, the limit to the pure diffusive result, and interesting effects of specific physical forms of the initial distribution of the particle. The latter finding provides, in principle, a method to extract information about the initial particle placement through short-time features of the time dependence of the survival probability.

Arbitrary relative locations of trap, potential center and particle at the initial moment result in consequences that are physically more interesting but necessitate numerical procedures of analysis. We have developed two such procedures, one based on a discretization method applied to the partial differential equation and the other based on a numerical Laplace transform (direct and inverse). We have studied the differential equation discretization method in the Appendix and shown, through comparison to analytic results, that it is  usable in all situations in which the Smoluchowski width is not too small relative to characteristic distances in the problem. Then, through a comparison of the discretization method to the Laplace method we have shown to latter to be reliable and have used it throughout the rest of the paper. 

Our results have uncovered noteworthy consequences of a potential: its introduction can help reaction (enhance capture) but also can hinder it if the potential pull is large enough. Section 4, in particular Figs. \ref{fig:Figure3} and \ref{fig:Figure4}, shows the results of our systematic study with symmetrical and non-symmetrical placement of trap and particle relative to the attractive center. Numerically observed features are satisfactorily explained in terms of uphill versus downhill motion of the particle in the potential and also by a study of where the trap lies relative to the equilibrium Smoluchowski distribution.
Two additional methods of analysis are provided in Section 5: one based on a transfer rate developed earlier by one of the  present authors in a different context and an approximate analysis for small capture based on the equilibrium Smoluchowski distribution. The former is represented by Fig. \ref{fig:Figure5} and Eqs. (\ref{k}), (\ref{kTrapCenter_imper}), and the latter by Fig. \ref{fig:Figure6} and Eq. (\ref{approxdqdt}). Non-monotonic behavior of capture efficiency as a function of potential steepness is seen in both cases. It is particularly transparent in Eq. (\ref{approxdqdt}): the steady state probability distribution, to which the rate of the survival probability is proportional, is clearly non-monotonic as a function of the  Smoluchowski width $\sigma=\sqrt{2D/\gamma}$.

We believe that a systematic reaction diffusion theory based on the Smoluchowski equation has not been provided earlier in the literature. Applications that we envisage are, as we have explained in the Introduction, to a variety of microscopic and macroscopic systems and phenomena including  artificial photosynthetic machines and enzyme ligand binding, and animal behavior and transmission of infection in the spread of epidemics.

\acknowledgments
This work was supported in part by the Consortium of the Americas for Interdisciplinary Science, by the Program in Interdisciplinary Biological and Biomedical Sciences of the University of New Mexico and by the University of Colorado, Colorado Springs BioFrontiers Center. 
\newpage
\appendix
\counterwithin{figure}{section}
\section{Accuracy of Numerics Employed}
\begin{center}
\begin{figure}[h!]
\includegraphics[scale=0.5]{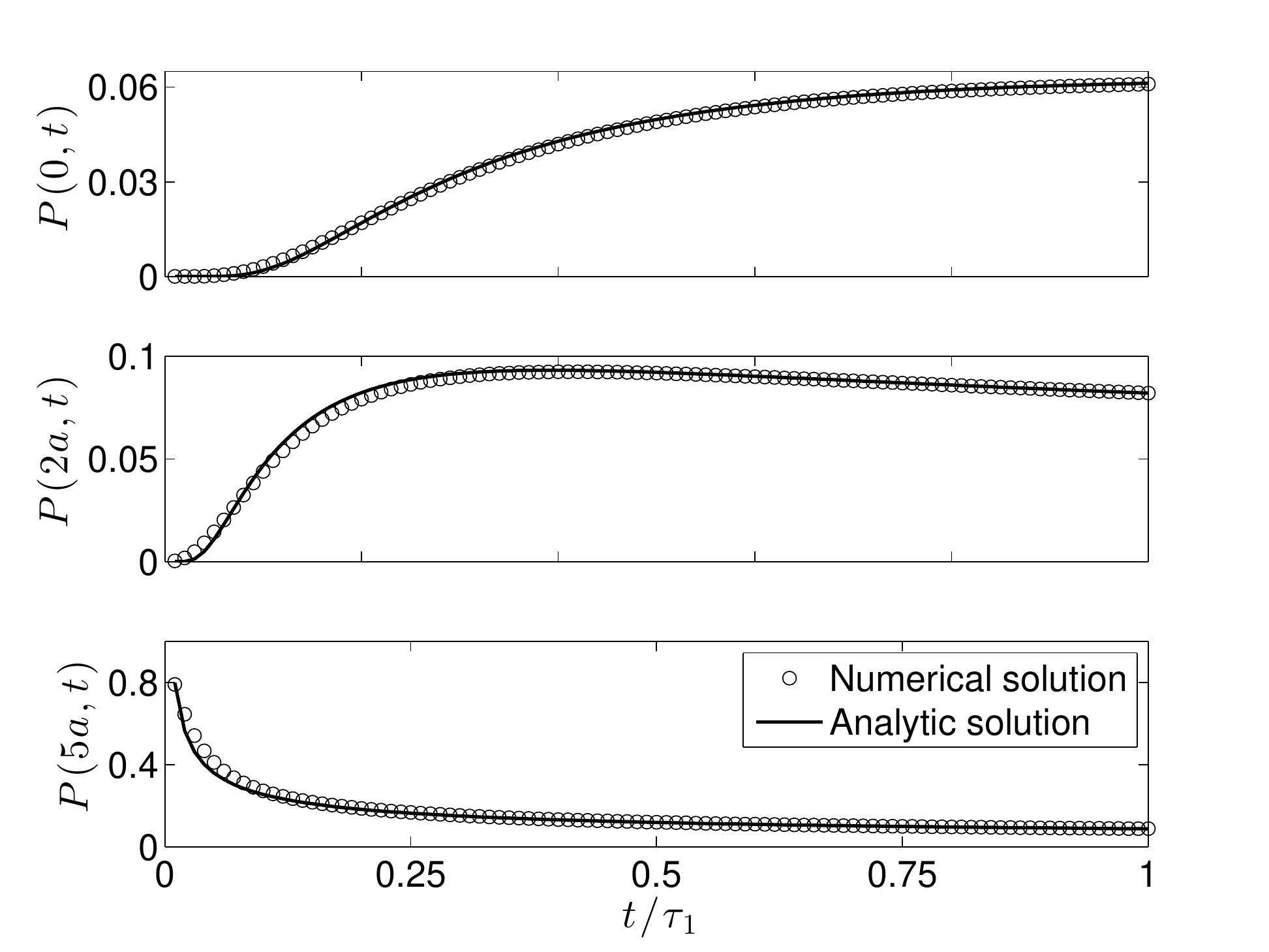}
\label{fig:AP1}
\caption{Comparison of the analytic and numerical solutions for the time evolution of probability density to find the walker at $x$ = 0, 2, and 5 in units of the lattice constant $a$. The initial condition is $P(x,0)=\delta(x-5a)$, and $f/F$ = 0.01. The total number of lattice points is 2001. Time is plotted in units of $\tau_{1}$.}
\end{figure}
\end{center}
In the following we demonstrate the extent of accuracy along with its limits, of one of the numerical methods we employ in the present paper, viz. the direct solution of the partial differential equation through discretization. Our purpose is to show details of the method, its success in the simplest case of the Smoluchowski equation without capture, i.e., Eq. (\ref{Smoluchowski_notrap}), and the range of parameters where its accuracy begins to drop. Below, we compare the results of the numerical procedure with exact (analytic) expressions for this capture-less case. This comparison provides confidence in the accuracy of the procedure as well as information about when not to use it (when the Smoluchowski width is too narrow). On the basis of this information we use  confidently (in sections 3 and 4) both the analytic Whittaker function expression derived in Eq. (\ref{centertrapW}) and \emph{another} semi-numerical  consisting of Laplace inversion of analytic expressions we obtain. 
\begin{center}
\begin{figure}[h!]
\includegraphics[scale=0.5]{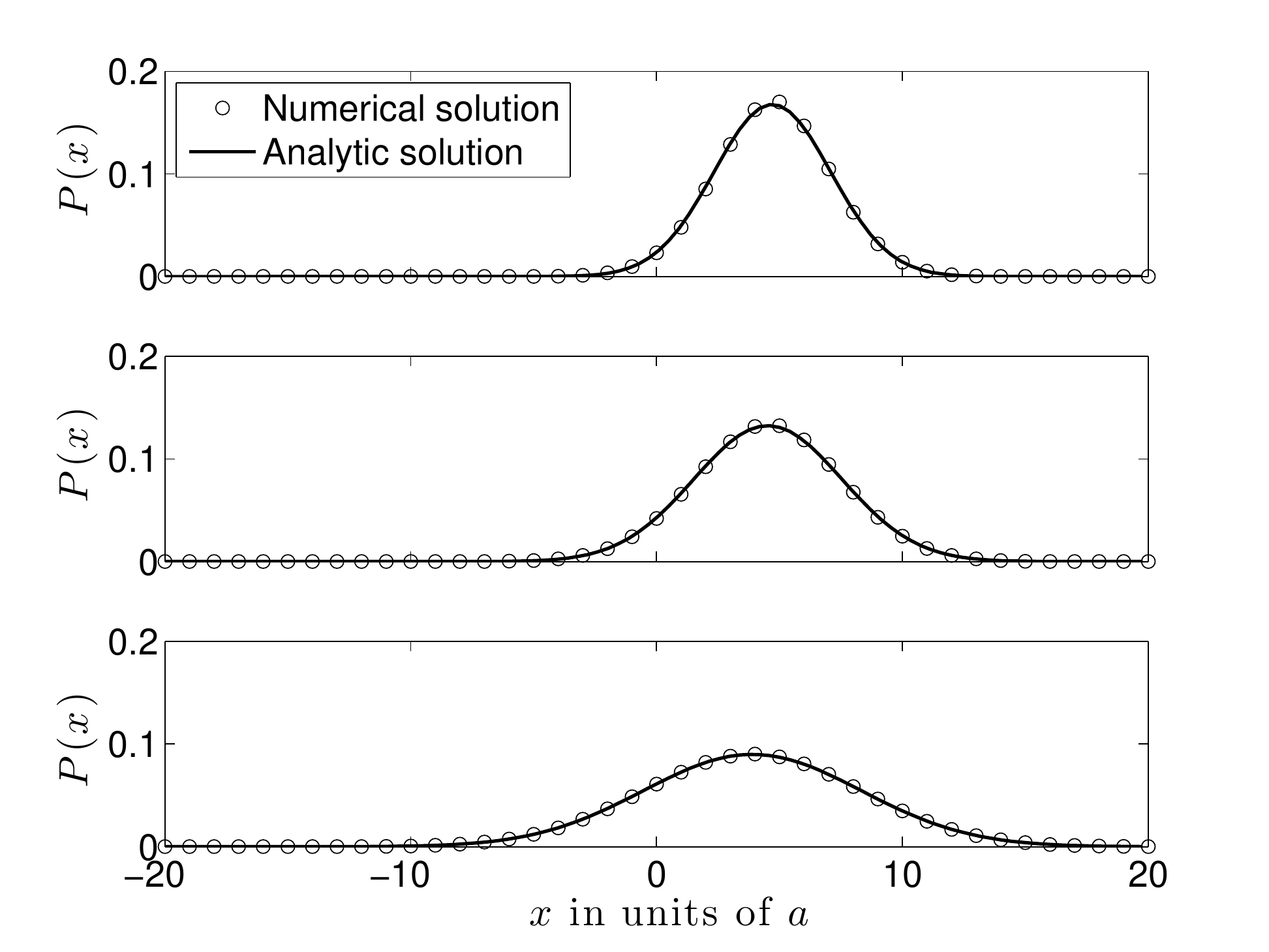}
\label{fig:AP2}
\caption{Comparison of the analytic and numerical solutions for the spatial distribution of the probility density at three different times. Agreement is excellent. The number of lattice points, and $f/F$ as in Fig. A1. The value of $t/\tau_{1}$ in the three panels from top down is 0.24, 0.40, and 1, respectively.}
\end{figure}
\end{center}

Equation (\ref{Smoluchowski_notrap}) can be looked upon as the continuum limit of the discrete equation
\begin{align}
 \label{discreteSmoluchowski}
 \frac{dP_{m}(t)}{dt}= &F\left[P_{m+1}+P_{m-1}-2P_{m}\right]\\
 &+ f\left[(m+1)P_{m+1}-(m-1)P_{m-1}\right]. \nonumber
\end{align}
Here $P_{m}(t)$ is the probability of occupation of a site $m$ at time $t$, $F$ is the nearest-neighbor hopping rate, $f$ is the rate of motion due to the attraction towards the center which is at $m=0$. If $a$ is the intersite distance, the continuum limit that transforms the above equation into the Smoluchowski Eq. (\ref{Smoluchowski_notrap}), is
\begin{align}
 \nonumber
  & a\rightarrow 0,\, F\rightarrow \infty,\, Fa^2\rightarrow D,\,f=\gamma/2,\\
 \nonumber
  & ma\rightarrow x,\, P_m(t)/a\rightarrow P(x,t).
 \end{align}

Therefore, if we write $m_0=x_0/a$, the combination of the above correspondence with the analytic propagator expression Eq. (\ref{prop}) gives
\begin{equation}
 a\Pi(x,x_{0},t) =
 \sqrt{\frac{f/F}{\pi(1-e^{-4ft})}}e^{-\frac{(m-e^{-2ft}m_{0})^{2}}{(1-e^{
-4ft})}}=  \Pi_{m,m_{0}}(t). \nonumber
\end{equation}
We use the middle expression above as the consequence of the analytic Smoluchoski equation in the discretized context, find $\Pi_{m,m_{0}}(t)$ numerically for a given $m_0$ and $f/F$ by using standard Matlab procedures such as Ode45, and compare the two to ascertain the extent of the accuracy of the numerical procedure.
\begin{center}
\begin{figure}[h!]
\includegraphics[scale=0.5]{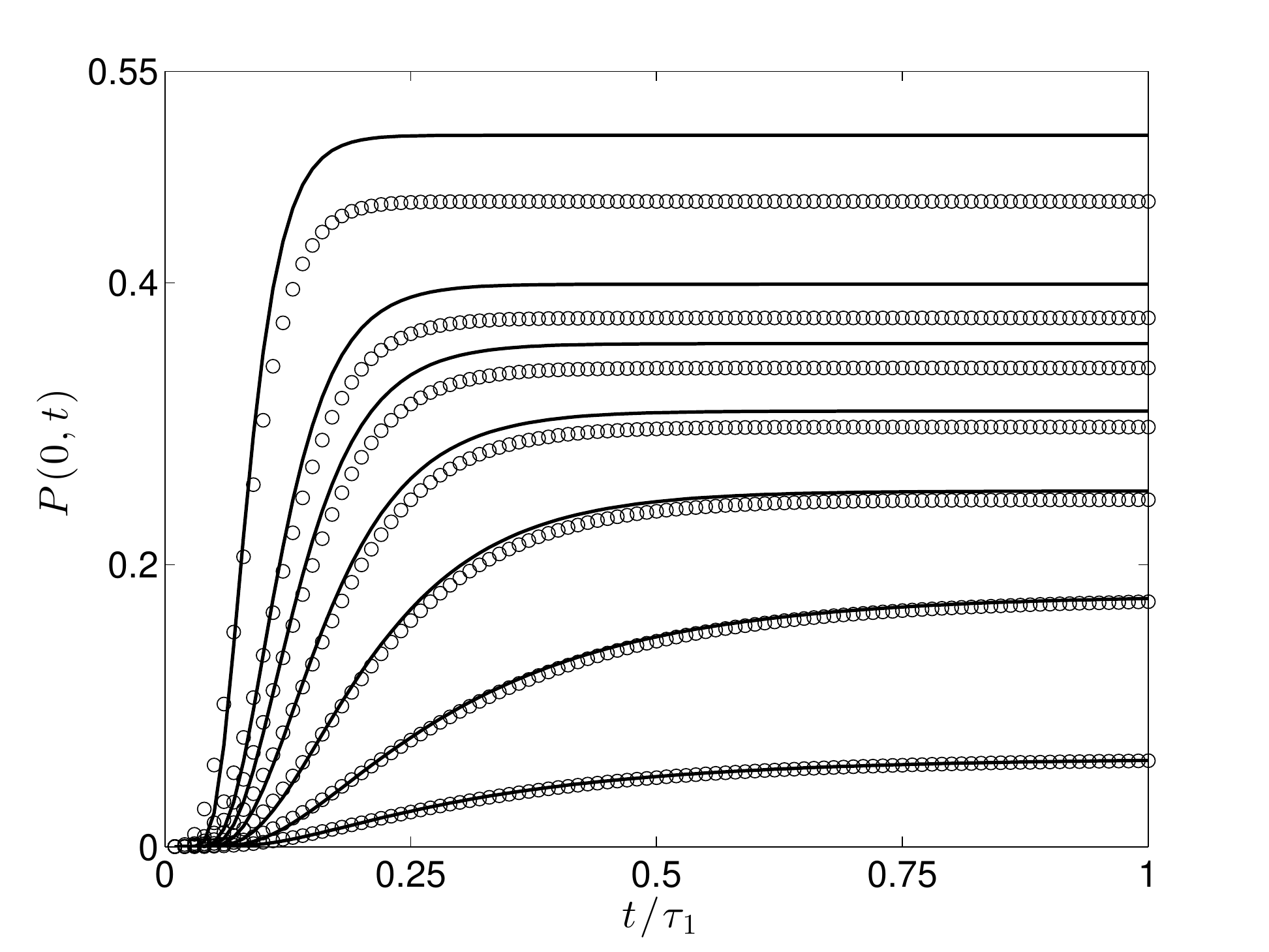}
\label{fig:AP3}
\caption{Inaccuracies in the discretization procedure that develop as the Smoluchoski width becomes smaller. Shown is the time evolution of the probability density at $x$ = 0 for various values of $f/F$. From the lower curve to the top, $f/F$ vaires as $0.01, 0.1, 0.2, 0.3, 0.4, 0.5, 08$. The circles and the solid lines depict the numerical and analytic solutions, respectively. Other parameters are as in Figs. A1 and A2. Noticeable departures are visible in the top four curves.}
\end{figure}
\end{center}

The result of the comparison for the initial condition at  $m_0$ = 10, for a total number of lattice sites equal to 2001 with periodic boundary conditions with $f/F=0.1$ is shown in Figs. A1 and A2 where time is plotted in units of $Ft$. In Fig. A1 we show the time evolution of the probability density  at three specific locations and in Fig. A2 the spatial distribution of the probability density at three specific times. Circles depict the numerical solution while lines are the analytic solution. Agreement is excellent and  demonstrates the generally satisfactory nature of the numerical procedure. 

The accuracy of our numerical procedure relies on the ratio $f/F$ being small. This quantity is inversely proportional to the square of the equilibrium width of the Smoluchowski dynamics; in discrete space, this width should not be smaller than the lattice constant $a$ of the discretization.  The numerical method based on discretization of the differential equation will thus begin to give inaccurate results when $a/\sigma = \sqrt{f/F}$ begins to get large. To demonstrate this limit of accuracy, we show Fig. A3. Our numerical solution starts to show deviation from the analytic solution when $f/F$ exceeds 0.15. Figure A3 shows the  evolution of the probability density at site $m$ = 0 for different values of $f/F$.

We conclude, first that for sufficiently small $f/F$ the discretization of the differential equation is a reasonably accurate procedure (because it reproduces well the analytic results), second that the analytic expression involving the Whittaker function, Eq. (\ref{centertrapW}), is accurate as its consequences agree with those of the present procedure (see section 3) provided the latter is used with small $f/F$, and finally that for large $f/F$ it is best to use the Laplace inversion numerical procedure as the discretization fails.



\begin{thebibliography}{99}
\bibitem{taunote}
The definition of $\tau_1$ we use in the present paper differs, by a factor of two, from that used in a recent article \cite{KSinJPC} by two of the present authors. There the definition had been  $\tau_1=x_0^2/D.$
\bibitem{width}
Standard usage would call the Gaussian width to be $\sqrt{D/\gamma}$. We have found it convenient to differ in our definition of $\sigma$ through a factor of $\sqrt{2}$.
\end{thebibliography}
\end{document}